\documentclass[12pt]{article}
\usepackage[english]{babel}
\usepackage{epsfig,sint,macros,macros_static}
\begin{document}

\begin{titlepage}
\begin{flushright}
  DESY 01-095 \\
\end{flushright}

\vskip 1 cm
\begin{center}
  {\Large\bf The $\nf=0$ heavy quark potential from short to intermediate distances}
\end{center}
\vskip 1.0cm
\begin{center}
{\large Silvia Necco and Rainer Sommer}
\vskip 0.8cm
DESY, Platanenallee 6, D-15738 Zeuthen, Germany
\vskip 0.8cm
 Silvia.Necco@desy.de
\vskip 0.1cm
 Rainer.Sommer@desy.de
\end{center}
\vskip 2.5ex
{\bf Abstract}
\vskip 0.7ex

We study the potential of a static quark anti-quark pair in the
range $0.05\,\fm \leq r \leq 0.8\fm$, employing a sequence of
lattices up to $64^4$.
Lattice artifacts in potential and force are investigated theoretically
as well as numerically and continuum quantities are obtained by extrapolation
of the results at finite lattice spacing.
Consistency of the numerical results
with the form of scaling violations predicted by an analysis 
\`a la Symanzik is found. The scale $\rnod/a$ is determined for the
Wilson action up to $\beta=6.92$.

  \vfill

\begin{flushleft}
  DESY 01-095\\
  July 2001
\end{flushleft} 
\eject

\end{titlepage}


\section{Introduction}
The $\Lambda$-parameter of quenched QCD has been computed \cite{mbar:pap1}
by means of an iterative finite size scaling method which allows
to connect long distance non-perturbative quantities with 
short distance renormalized couplings in a controlled manner 
\cite{alpha:sigma}. The result is the $\Lambda$-parameter in units
of the low energy scale $\rnod$. The latter is equivalent to the force
between static quarks at intermediate distances $r=\rnod\approx0.5\fm$
\cite{pot:r0}.

Given this information, the perturbative expansion for the
static potential 
\cite{Fischler:1977yf-Schroder:1998vy}
becomes a {\em parameter free} prediction for the 
force between static quarks at short distances.

On the other hand, the potential itself 
was the very first quantity 
to be computed  by
Monte Carlo simulation of the pure gauge theory \cite{pot:creutz} 
and it is
therefore very interesting to compare perturbation theory 
with the non-perturbative force.
Since the original work of Creutz in the SU(2) gauge theory,
many computations have been performed 
in SU(3) 
\cite{pot:rebbi2-pot:hartmut}
but the short distance region has not yet 
been investigated convincingly. The reason is simple:
by ``the potential'' we do of course mean the potential
in infinite volume; in practice we know that for the force
at distances
up to $r \approx 0.5\,\fm$, the deviation from
infinite volume are small on an $L^4$ torus with $L=1.5\,\fm$ 
\cite{pot:r0_SU3}.  With such a value of $L$, an
investigation of distances of $r\approx 0.05\,\fm$ 
with control over $\rmO(a/r)$ lattice artifacts requires very 
large lattices. 

In the $\nf=0$ theory (for our purposes this means
in the pure Yang Mills theory) large lattices may 
nowadays be simulated. Still it would be very costly
to perform a computation of the force in the full range 
say $0.05\,\fm \leq r \leq 0.8\fm$, since large distances on 
fine lattices require also large time extents of the
Wilson loops and -- with presently known
techniques -- compute intensive smearing methods.

To investigate this range we therefore
separately consider two regions, $0.05\,\fm \leq r \leq 0.3\fm$
and $0.2\,\fm \leq r \leq 0.8\fm$, where the first one
needs very large lattices and small lattice spacings 
and the second one had been simulated before on
coarser lattices \cite{pot:r0_SU3}. The region of overlap
serves for calibration.
Our reference scale, $\rnod$, is computed in the second range
and related to $\rc \approx 0.26 \fm$, which is accessible 
in both.

Apart from the wish to test renormalized perturbation theory,
an additional motivation for the computation presented here
is that we can 
enlarge the region of bare couplings $\beta$ for which
$\rnod/a$ is known. It is
now extended up to $\beta=6.92$. 
This opens up the possibility to
investigate scaling violations
over a larger range of lattice spacings, which is
of interest in view of the unexpected results
found in the 2-d sigma model \cite{sigma:unexpected}.
There indications for scaling violations 
differing from their form  expected from
an analysis \`a la Symanzik have been found.
In our paper we shall first apply Symanzik's theory of
lattice spacing effects
to the static potential, closing a gap in the literature. 
We then discuss our numerical results for $a$-effects.
A detailed comparison of the non-perturbative force and
potential with perturbation theory will be presented in a separate
publication.


\section{The scales $\rnod$, $\rc$ \label{s_scales}}
\subsection{Definition}
In a pure gauge theory we have to specify one
physical quantity in order to renormalize the theory. 
While a renormalized coupling
in a non-perturbative scheme may be used, it is more convenient to choose
a dimensionful long distance observable, which introduces {\em the} scale into the 
theory. Predictions of all other dimensionful quantities are then expressed
in units of this scale. They are well defined and may be extrapolated to the 
continuum limit from results at finite resolution. It is clear that the
scale should be chosen with care since it influences the precision
of many predictions. The length scale $\rnod$, defined 
in terms of the force $F(r)$ between static quarks by
the implicit equation \cite{pot:r0}
\be\label{e_rnod}         
r^{2}F(r)\big|_{r=r(c)}=c\,, \quad \rnod=r(1.65)\,,
\ee
has turned out to be a good choice: it may be computed with
good statistical and systematic precision.
In QCD, $\rnod$ has a value of about 0.5~fm  \cite{pot:r0}. 

As discussed in the introduction, it is convenient to choose a 
smaller reference length scale when one is interested
in short distance properties of the theory. We therefore
introduce also
\be  \label{e_rc}
  \rc=r(0.65)\,
\ee
and would like to know its relation to $\rnod$.

\begin{figure}
\begin{center}
\includegraphics[width=8cm]{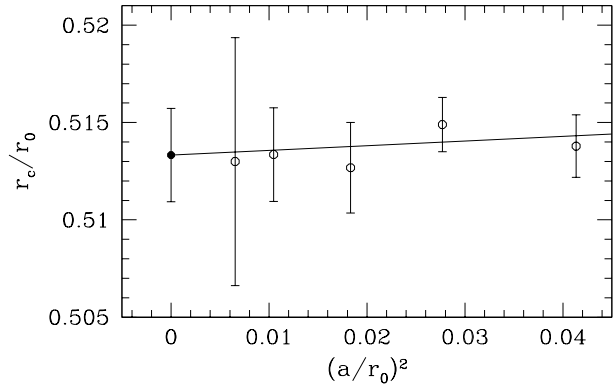}
\caption{\footnotesize{The ratio $\rc/\rnod$ for 
$5.95\leq\beta\leq 6.57$ in the standard Wilson action (circles) including
the continuum extrapolation (solid point). 
This is an analysis of data in \protect\cite{pot:r0_SU3}. 
}\label{f_rcrnod}}
\end{center}
\end{figure}

\subsection{The ratio $\rc/\rnod$}
The evaluation of $r(c)$ from \eq{e_rnod} begins with the extraction of the 
static quark potential $V(r)$ between quarks separated 
by a distance $r$ along a lattice axis. We use Wilson 
loop correlation functions from \cite{pot:r0_SU3}
as well as new simulations for smaller values of the lattice 
spacing ($\beta=6.57,\,6.69,\,6.81,\,6.92$) keeping the
size of the $L^4$-torus at $L \approx 3.3 \rnod$,
where finite size effects in $V(r)$ are known to be small for
$r\leq\rnod$.
For the smallest lattice spacing this required a $64^4$ lattice.
We essentially followed the procedure of \cite{pot:r0_SU3}.
Some details are given in \app{app_a}. 

The force at finite lattice spacing is 
then defined as
\be\label{e_force}
F(\rI)= \left[V(r)-V(r-a) \right]/a,
\ee
with the distance $\rI$ chosen such that
the force on the lattice has no deviations from the force in the 
continuum when evaluated at tree level. 
Lattice artifacts are then suppressed by a power of $\alpha$;
$F(\rI)$ is a tree-level improved observable. 
In the SU(2) theory it has been observed that the remaining
lattice artifacts in the force are surprisingly small
\cite{pot:r0} and here we shall again find no evidence for them 
as long as $r>2a$.

Explicitly we have 
\be
\label{e_rI}
(4\pi \rI^2)^{-1}=\left[G(r,0,0)-G(r-a,0,0)\right]/a,
\ee
where $G(\bf{r})$ is the (scalar) lattice propagator in 3 dimensions
defined in \eq{green}. 
With the methods of \cite{coord_space}
it can be constructed directly in 
coordinate space. To our knowledge these methods have not been 
applied in 3-d and we had to find the necessary invariant
of the recursion relation \eq{recursion}.
Formulae are listed  in \app{app_d}. 

Solving \eq{e_rnod} requires furthermore an interpolation of $F(\rI)$.
This
can be done with small systematic errors \cite{pot:r0}, which nevertheless
dominate over the statistical errors at small $r$ (see \app{app_a} for
details).

\Fig{f_rcrnod} shows the ratio $\rc/\rnod$ for several lattice spacings.
No dependence on the resolution
is seen within the errors of below 1\%. A
continuum extrapolation gives
\be\label{e_rcrnodcont}
{\rc/\rnod}=0.5133(24)
\ee
and we note that it is safe to use \eq{e_rcrnodcont} also at finite lattice 
spacings starting around $\beta=6.4$.

\subsection{Parameterization of $\rnod/a$}

\begin{figure}
\begin{center}
\includegraphics[width=8cm]{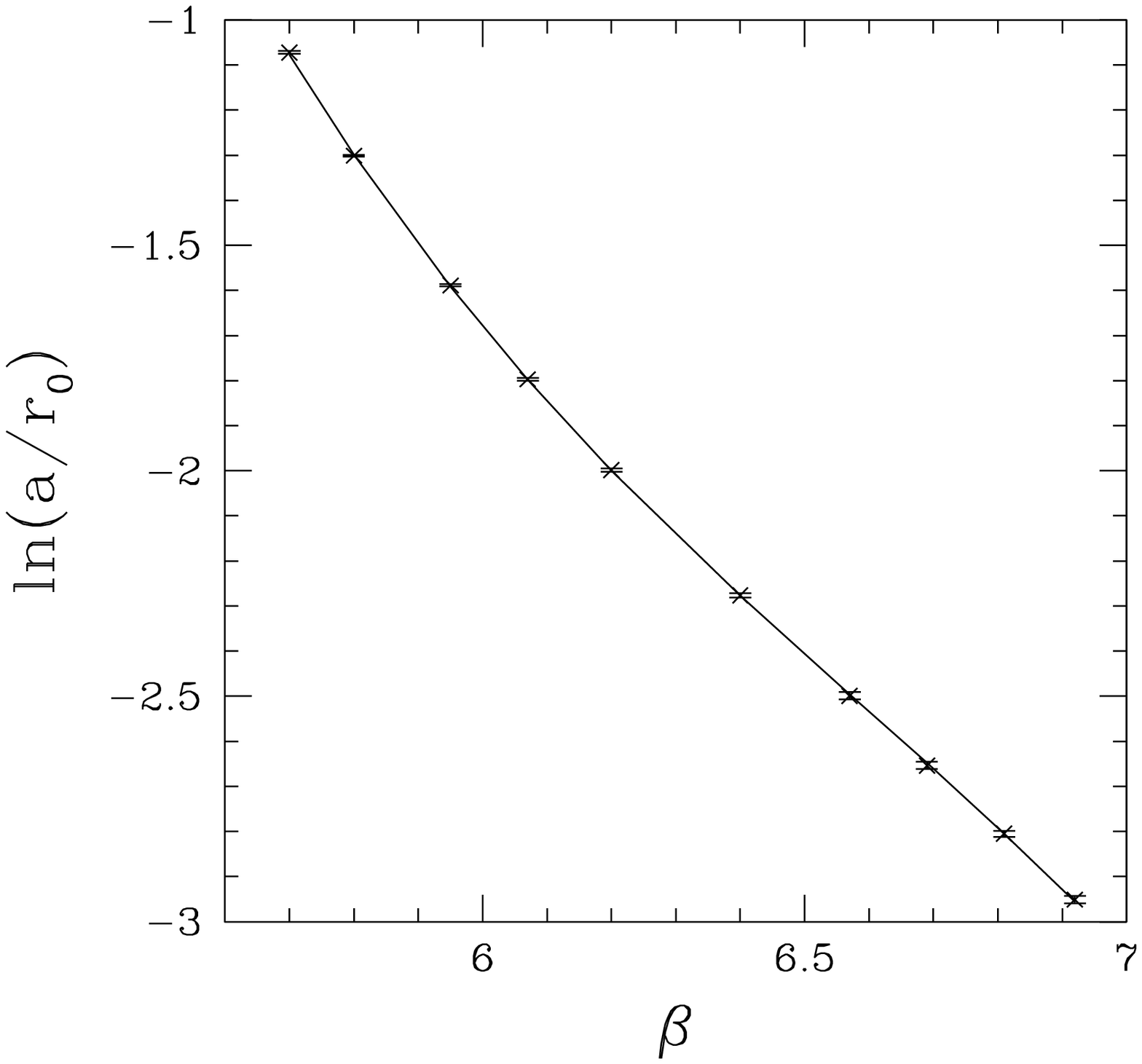}\caption{\footnotesize
Interpolation of $\rnod/a$.
\label{f_fitrnod}}
\end{center}
\end{figure}

The direct determination of $\rnod/a$ for $5.7\leq\beta \leq 6.4$ \cite{pot:r0_SU3} \footnote{
In \cite{pot:r0_SU3} the simulation for $\beta=6.57$
was performed with a single smearing level, yielding less control
over excited state contaminations. As a result
the error on $\rnod/a$ seems to be somewhat underestimated. Although the
statistical significance of this
small effect is not clear, we decided to use our new data for $\beta=6.57$
instead.}
and our new computations of $\rc/a$ in the range $6.57\leq\beta\leq 6.92$ may be combined 
with $\rc / \rnod$=0.5133(24) to obtain an interpolating formula giving
$\rnod/a$ in the whole range $5.7\leq\beta\leq 6.92$. Following \cite{pot:r0_SU3} 
we interpolate $\ln(a/\rnod)$ through a polynomial in $\beta$ and find that
\bes \label{e_rnodfit}
\ln(a/\rnod)&=&-1.6804-1.7331(\beta-6)+0.7849(\beta-6)^{2}-0.4428(\beta-6)^{3}\,, 
     \nonumber \\ 
  && {\rm for} \;\; 5.7\leq\beta\leq 6.92, 
\ees
is an excellent approximation (\fig{f_fitrnod}) to the MC results. 
This  formula covers a
larger range of $\beta$ than the one given in \cite{pot:r0_SU3},
but its precision in the low $\beta$ range is somewhat worse. 
The accuracy of 
$\rnod/a$ in 
\eq{e_rnodfit} is about $0.5\%$ at low $\beta$ decreasing to $1\%$ 
at $\beta=6.92$.

\begin{table}
\caption{\label{t_rnod}\footnotesize{Results for $\rnod/a$ \protect\cite{pot:r0_SU3}
  and $\rc/a$. }}
\begin{center}
\begin{tabular}{llcll}
\hline
$\beta$ & $\rnod/a$ & & $\beta$ & $\rc/a$\\
\hline
5.7 &  2.922(9) &~~~~~~& 6.57 & 6.25(4)\\
5.8 &  3.673(5) &~~~~~~& 6.69 & 7.29(5) \\
5.95 & 4.898(12)&~~~~~~& 6.81 & 8.49(5)\\
6.07 & 6.033(17)&~~~~~~& 6.92 & 9.82(6) \\
6.2 & 7.380(26)  \\
6.4 & 9.74(5)   \\
\hline
\end{tabular}
\end{center}
\end{table}


\section{Continuum force and potential}

We now determine the continuum force and potential by
extrapolation of the MC-results at finite values of
the lattice spacing to the continuum. It is expected that 
close to the continuum limit
the dominant discretization error is quadratic
in the lattice spacing. A clean argument why this is so 
has never been given in the literature. We shall fill
this gap and give one in the next section. 

\begin{figure}
\begin{center}
\includegraphics[width=10.5cm]{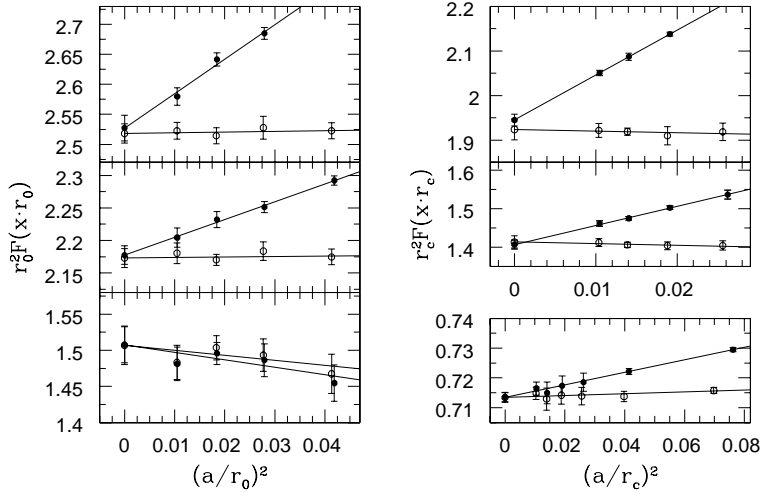}
\vspace{-0.5cm}
\caption{\footnotesize
Continuum
extrapolation of $\rc^2 F(x\rc)$, for $x=0.4,~0.5,~0.9$ from top to bottom
and of $\rnod^2 F(x\rnod)$, for $x=0.5,~0.6,~1.5$ from top to bottom. 
The data are from our new computations and from \protect\cite{pot:r0_SU3}. 
Filled circles correspond to the naive value $\rI=r-\frac{a}{2}$ instead
of \protect\eq{e_rI}.
\label{f_force_extr}}
\end{center}
\end{figure}

\Fig{f_force_extr}  shows the dependence of the force in
units of $\rc$ on the resolution
for some selected values of the separation $r$. Fitting 
\bes  \label{e_force_extr}
  \rc^2 F(r) = \left.\rc^2 F(r) \right|_{a = 0}\left[1 + s \times (a/\rc)^2\right]\,
\ees
for fixed $r/\rc$, the slope $s$ is statistically not significant throughout
our range of $r$ and $\beta$. Our statistical precision
allows to quote
\bes \label{e_slope}
 |s| < 1\; {\rm for}\; r/a > 2\,, \; 0.4\leq r/\rc  \leq 1\,. 
\ees
Of course this 1-$\sigma$ bound is valid only when all 
details are as discussed above, in particular eqs.~(\ref{e_force},\ref{e_rI})
are used to define the force at finite $a$. If the naive form
$\rI=r-\frac{a}{2}$ is employed instead, the 
corresponding slopes $s$ become rather large, as can be seen in the figure. 

A similar statement,
\bes
   \rnod^2 F(r) &=& \left.\rnod^2 F(r) \right|_{a = 0}
                    \left[1 + s \times (a/\rnod)^2\right]\,,  \\
  |s| &<& 1.2\; {\rm for}\; r/a > 2\,,  \; 0.5\leq r/\rnod  \leq 1.5\,, 
\ees
can be made for larger $r$. 

\begin{figure}
\begin{center}
\includegraphics[width=9cm]{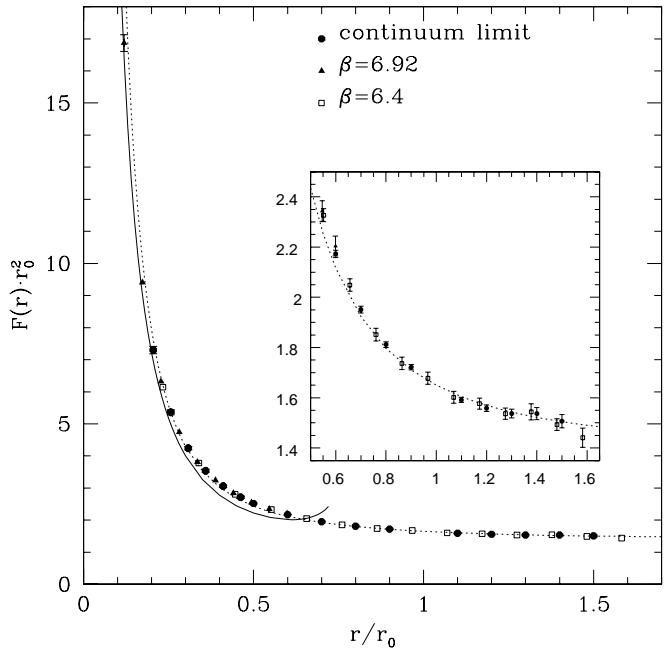}
\caption{\footnotesize{The force in the continuum
limit and for finite resolution, where the discretization
errors are estimated to be
smaller than the statistical errors. The full line
is the perturbative prediction with $ \Lambda_\msbar\, \rnod =0.602 $. 
The dashed curve corresponds to the bosonic string model normalized by 
$\rnod^2 F(\rnod)=1.65$.}\label{f_force}}
\end{center}
\end{figure}

The continuum
force is plotted in \fig{f_force} using \eq{e_rcrnodcont} to 
combine the two regimes of $r$.
Some data at finite $\beta$ are included in the figure. In these
cases we used our ``bounds'' on $s$ to estimate that the discretization 
errors are smaller than the statistical ones.

For large values of $r$, the force is expected to be given by a
constant, the string tension, plus a first universal \cite{pot:luescherterm} 
$1/r^2$ correction. Assuming this description to be valid already
at $r=\rnod$ yields the parameter free bosonic
string model,
\bes \label{e_string}
F(r)=\sigma+{\pi \over 12r^{2}}\;   , \;\; \sigma\rnod^2=1.65-\pi/12\,,
\ees
which is in excellent agreement with our results for $r \geq 0.8\rnod$. 
Note that in the same region of $r$, excited potentials do not at all follow
the expectations from an effective bosonic string theory \cite{pot:juge}.
This suggests that the agreement with \eq{e_string} is rather 
accidental.
In any case one would expect corrections to
this formula to be negligible only for much larger
$r$.
Nevertheless \eq{e_string} is a very good effective 
description of $F(r)$
for $0.8\rnod \leq r \leq 1.6 \rnod$. 

At short distances the force may be obtained
by an integration of the 
perturbative renormalization group, 
\bes
F(r) &=& C_{\rm F} \gqq^{2}(r)/(4\pi r^2)\,, \quad C_{\rm F}=4/3\,,\\
-r {\rmd \over \rmd r} \gqq &=& \beta(\gqq)= -\sum_{\nu=0}^{2}b_{\nu}\gqq^{2\nu+3}\,,\label{pert} \\
 b_0&=& \frac{11}{16\pi^2}\,,\; b_1=\frac{102}{(16\pi^2)^2}\,,\;
b_2= \frac{1}{(4\pi)^6}\left(-3470 + 2519\frac{\pi^2}{3}  - 
99\frac{\pi^4}{4}  + 726\zeta(3)\right) \nonumber
\ees
Here the 3-loop coefficient, $b_2$, could be extracted from 
\cite{Schroder:1998vy,Peter:1997me,Melles:2000dq}.
Inserting \cite{mbar:pap1}
\be \label{e_Lambdarnod}
 \Lambda_\msbar\, \rnod =0.602(48)
\ee
as well as the known relation between $\Lambda_\msbar$ and $\Lambda_{\rm qq}$
\cite{Fischler:1977yf,Billoire:1980ih}
to fix the integration constant, 
we have a parameter free perturbative prediction. It agrees
with the non-perturbative results up to $r\approx 0.3\rnod$. 
Indeed, inserting $\Lambda_\msbar\, \rnod$ at the upper end
of the error bar of \eq{e_Lambdarnod} into the
perturbative formula, very close agreement with the 
data points is seen in this range of $r$.
Again this agreement at quite
large $r$ appears somewhat accidental as the perturbative
prediction itself is only stable at smaller distances. 
Certainly there is no need for large non-perturbative terms 
at such distances as it was concluded earlier 
on the basis of an exploratory investigation \cite{pot:bali99}.
We shall discuss the comparison with perturbation theory 
in more detail in a separate publication. 

\begin{figure}
\begin{center}
\includegraphics[width=11cm]{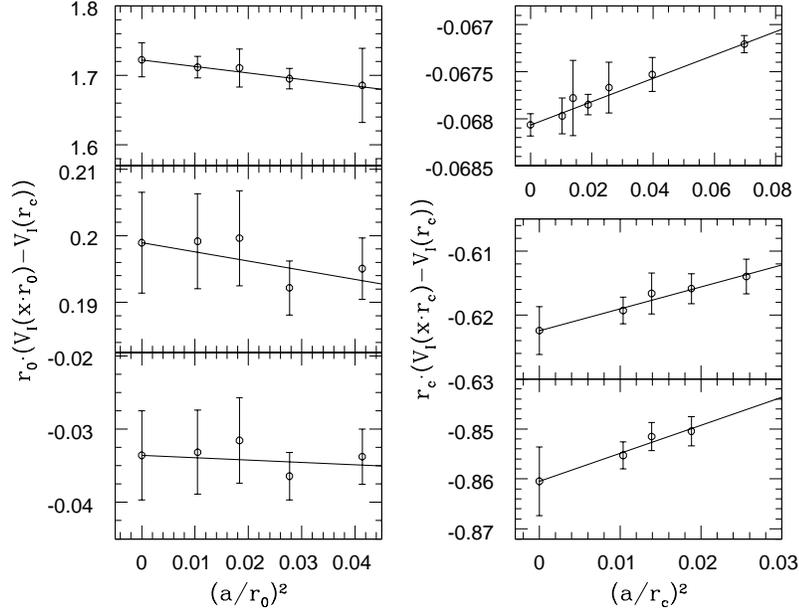}\caption{\footnotesize 
Continuum extrapolation 
of $\VI(r)$, for $r/\rnod=1.5,\,0.6,\,0.5$ on the left hand side and
for $r/\rc=0.9,\,0.4,\,0.3$ on the right hand side.
\label{f_pot_extr}}
\end{center}
\end{figure}
The potential contains the same physical 
information as the force but statistical and systematic errors 
in the lattice determination
are different. To eliminate the self energy contribution
we consider $\VI(r)-\VI(\rc)$, where we apply a tree-level improvement 
as for the force,
\be\label{pot_impr}
\VI(\rIv)=V(d)\,\, ,\,\, (4\pi\rIv)^{-1}= G(d,0,0).
\ee 
The continuum extrapolation is performed exactly as for the 
force. In this case slopes, $s$, defined as above are statistically
significant for both our smallest distances ($r\approx 0.3\rc$) and the
largest one ($r\approx 1.5\rnod$). As a consequence, at small $r$ the 
combination $[\VI(r)-\VI(\rc)]\rc$
has about $1\%$ discretization errors at $\beta=6.92$ while
in the large distance region these errors go up to $1\%$ in
$[\VI(r)-\VI(\rc)] \rnod$ at $\beta=6.4$. 
These are the $\beta$-values corresponding to
the smallest lattice spacings available in the 
two different regions.

\begin{figure}
\begin{center}
\includegraphics[width=9cm]{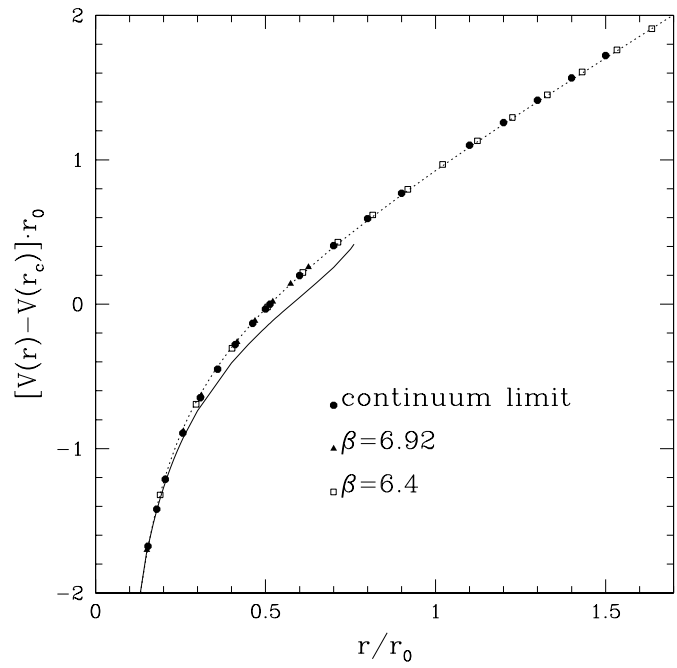}
\caption{\footnotesize{The static potential. 
The dashed line represents the bosonic string model 
and the solid line the prediction of perturbation theory
as detailed in the text.}\label{v_pot}}
\end{center}
\end{figure}

The continuum potential is plotted in \fig{v_pot}, with some data at finite 
$\beta$. In this figure the statistical as well as the discretization errors
are below the size of the symbols. 
In the whole range of $r$ the non-perturbative 
results are described by the model \eq{e_string} 
within about  1\% accuracy. For short distances, $r<0.3 \rnod$,
the perturbative prediction $V(r) = V(0.3 \rc)+\int_{0.3\rc}^r \rmd y F(y)$
(with the perturbative expansion for $F$ as discussed above) is quite accurate.
For future reference continuum force and potential are listed in
\tab{t_cont}.
  \begin{table}[ht] 
\caption{\label{t_cont}\footnotesize{Potential and force after 
continuum extrapolation. }}
\begin{center}
  \begin{tabular}{ c  c  c  r  c  c}
  \hline 
$r/\rc$ & $\rnod^2 F(r)$ &  $\rnod(V_{\rm I}(r)-V_{\rm I}(\rc))$  &~~~~~~ $r/\rnod$ & $\rnod^2 F(r)$ &  $\rnod(V_{\rm I}(r)-V_{\rm I}(\rc))$\\ 
  \hline 
 $ 0.3  $ & $   $ & $ -1.676 ( 16 )$ & $ 0.5  $ & $ 2.518 ( 16 ) $ & $ -0.0336 ( 61 )$ \\ 
 $ 0.4 $ & $ 7.30 ( 11 ) $ & $ -1.2125 ( 93 )$ & $ 0.6  $ & $ 2.173 ( 15 ) $ & $ 0.1989 ( 76 )$ \\
 $ 0.5 $ & $ 5.363 ( 78 ) $ & $ -0.8926 ( 74 )$ & $ 0.7  $ & $ 1.951 ( 14 ) $ & $ 0.4051 ( 89 )$ \\
 $ 0.6 $ & $ 4.244 ( 53 ) $ & $ -0.6475 ( 54 )$ & $ 0.8  $ & $ 1.812 ( 11 ) $ & $ 0.5930 ( 99 )$ \\
 $ 0.7 $ & $ 3.538 ( 48 ) $ & $ -0.4494 ( 35 )$ & $ 0.9  $ & $ 1.722 ( 11 ) $ & $ 0.769 ( 11 )$ \\
 $ 0.8 $ & $ 3.060 ( 38 ) $ & $ -0.27950 ( 15 )$ & $ 1.1  $ & $ 1.592 ( 10 ) $ & $ 1.101 ( 11 )$ \\
 $ 0.9 $ & $ 2.713 ( 30 ) $ & $ -0.13259 ( 67 )$ & $ 1.2  $ & $ 1.559 ( 13 ) $ & $ 1.258 ( 12 )$ \\
 $     $ & $              $ & $                $ & $ 1.3  $ & $ 1.537 ( 18 ) $ & $ 1.413 ( 13 )$ \\
 $     $ & $              $ & $                $ & $ 1.4  $ & $ 1.537 ( 24 ) $ & $ 1.567 ( 14 )$ \\
 $     $ & $              $ & $                $ & $ 1.5  $ & $ 1.507 ( 27 ) $ & $ 1.722 ( 24 )$ \\
  \hline 
  \end{tabular} 
\end{center}
  \end{table} 


\section{Lattice artefacts}

The standard framework for the discussion of lattice artefacts is Symanzik's
effective theory, which is expected to give the asymptotic expansion of
suitable lattice observables in integer powers of the lattice spacing 
(up to logarithmic modifications) in an asymptotically free theory
\cite{impr:Sym1,impr:Sym2}. In the 2-dimensional O(3) $\sigma$
model it predicts this expansion to start at order $a^2$ but
unexpectedly numerical results 
are described better by a dominant linear term in $a$ for a
range of lattice spacings \cite{sigma:unexpected}. 
While there is no evident contradiction
with the result of an analysis \`a la Symanzik which is supposed to
describe the {\em asymptotic} behavior, the standard picture
should be tested in 4-d gauge theories and QCD as much as possible.

Below we shall consider some
observables and study their $a$-dependence. 
Before coming to these numerical examples, we want to give a brief 
but thorough argument that the leading artefacts in potential
differences are expected to be $\rmO(a^2)$. This is not obvious
since the potential is usually defined in terms of a 
Wilson loop, which does not fall into the category of
correlation functions of local fields discussed by Symanzik.
 
\subsection{Why are the leading lattice spacing errors to $F(r)$ quadratic
            in $a$?}

In a short version, our answer to this question is that the heavy quark 
effective theory \cite{stat:eichhill1}
formulated on the lattice is $\Oa$-improved without
adding any additional operators to the Lagrangian \cite{zastat:pap1}.
The potential is an energy of the effective theory and thus 
has no linear $a$-effects. Of course, for this statement to be meaningful,
the self energy has to be eliminated by 
considering potential differences or the force.
We want to also mention that our argument is
based on the assumption that the effective theory is renormalized as 
usual by the addition of local operators. Explicit perturbative
computations support this assumption 
but it has not been proven so far.

We find it easiest to discuss the $a$-effects starting from 
a correlation function in the effective theory with \SF boundary
conditions \cite{SF:LNWW,SF:stefan1,zastat:pap1}. 
A correlation function is chosen which describes a
 quark-antiquark pair separated by  $\vecx$  in space and 
propagating in Euclidean time from
$x_0=0$ to $x_0=T$. Boundary
conditions are taken exactly as specified in the quoted papers ($C=C'=0$).
In the notation of \cite{zastat:pap1}, the correlation function is 
given by
\bes
  \khh(\vecx,T) = \left\langle 
               \zetahb({\bf0}) \gamma_5 \zeta_{\bar{\rm h}}({\vecx}) \; 
               {\zeta_{\bar{\rm h}}\kern1pt'}({\vecx}) \gamma_5 \zetahprime({\bf0})  
                  \right\rangle \,.
\ees
In the standard formulation of the lattice theory, a positive transfer matrix exists
\cite{Luscher:TM}
and therefore $\khh$ has the exact representation 
\bes
  \khh(\vecx,T) = \left\{ \sum_{m\geq0} B_m \rme^{-E_m T} \right\}^{-1}
                  \left\{ \sum_{n\geq0} C_n(\vecx) \rme^{-V_n(\vecx) T} \right\}   
\ees
with positive $B_m,C_n(\vecx)$. The energies $E_m$ are the (negative) logarithms
of the transfer matrix in the vacuum sector and  $V_n(\vecx)$ are the energies in
the proper charge sector (see e.g. \cite{Knechtli:1999tw} for a discussion
of the transfer matrix 
in the presence of static charges). 
Adopting the convention $E_0=0$ (vanishing vacuum energy), 
the potential is given by 
$V(r)=V_0(r,0,0)$.

Renormalization and improvement of $\khh(\vecx,T)$
follows directly
from the discussion of \cite{zastat:pap1}:
the renormalized and 
improved correlation function, \footnote{
        It is assumed here that the improvement coefficient $\ct$ is
        chosen properly to remove $\Oa$ effects specific to the \SF \cite{alpha:SU3}.
        When full QCD with quarks is considered, one furthermore 
        has to account for $\csw,\cttilde$ \cite{impr:pap1}.  
        In massive QCD, the term proportional to $\bstat$ has to be 
        included in \eq{e_khhR} but plays no r\^ole in the following.}
\bes \label{e_khhR}
  \khhR(\vecx,T) = \rme^{-2\dmstat T}\;  \zh^4\; \khh(\vecx,T)\,,
\ees 
satisfies
\bes \label{e_khhcont}
  \khhR(\vecx,T) = \left. \khhR(\vecx,T) \right|_{a=0} + \rmO(a^2)\,
\ees 
and the renormalization constants $\dmstat$ and $ \zh$ 
may be chosen to depend 
on nothing but the bare coupling.  From \eq{e_khhcont} we may conclude
\bes\label{e_Vcont}
  V(r_1)-V(r_2) = \left[V(r_1)-V(r_2) \right]_{a=0} + \rmO(a^2)\,,
\ees 
since 
\bes
  V(r) = V_0(r,0,0) = a^{-1} \lim_{T\to\infty} 
                    \ln {\khhR(\vecx,T-a) \over \khhR(\vecx,T)} 
\ees
and
\bes
\ln {\khhR(\vecx,T-a) \over \khhR(\vecx,T)} = 
     \ln {\khh(\vecx,T-a) \over \khh(\vecx,T)} +2a\dmstat\,.
\ees
Our result \eq{e_Vcont} is valid for the pure gauge theory as
well as for the $\Oa$-improved formulation of QCD discussed in
\cite{impr:SW,impr:pap1}.

\subsection{Lattice spacing effects in force and potential.}

Examples for $a$-effects are shown in \fig{f_force_extr} and
\fig{f_pot_extr}. The following observations are relevant.
      Using eqs.~(\ref{e_force},\ref{e_rI}), i.e. a tree-level
      improved definition of the force, lattice spacing effects
      are below the level of our small statistical errors when
      one restricts oneself to $r\geq2a$ and the lattice spacings
      considered here. 
      By contrast, with a naive midpoint rule for the force, the $a$-effects 
      are quite sizeable. The difference is by construction a 
      pure power series in $a$, starting with $a^2$.
      For $r\geq2a$ this series is well approximated by the leading term.
      Thus both data sets in \fig{f_force_extr} contain the same
      essential information: full compatibility with Symanzik's
      theory of discretization errors.
The case of finite potential differences, \fig{f_pot_extr}, 
differs only slightly. Here, small $a$-effects could also be observed
for the tree-level improved definition. Again -- but not shown in the 
figure -- the standard definition without tree-level improvement
shows quite large $a$-effects. This has been known for a long time
\cite{pot:rebbi1,pot:rotinv}. In the past, usually a correction term,
relying on a fit to the potential,  
has been applied to remove the dominant $a$-effects
\cite{pot:michael_SU2}.
As shown in \cite{pot:r0}, and again here, this procedure
may be replaced by a simple and theoretically sound definition.

\subsection{Continuum extrapolation of $\Lmax /\rnod$ \label{s_lmaxrnod}}

In \cite{mbar:pap1,pot:r0_SU3}, a calculation of the $\Lambda$-parameter
of the pure gauge theory was presented. The result may be split in
the following way,
\bes
\Lambda_\msbar r_0 =  (2.0487 \Lambda  \Lmax) / ( \Lmax/\rnod) \,.
\ees
Here $\Lmax$ is the distance  where the \SF coupling \cite{alpha:SU3} has
the value $\gbar^2(\Lmax)=3.48$ and the combination $\Lambda  \Lmax = 0.211(16)$ 
in the first parenthesis is obtained from the 
non-perturbative evolution of the \SF coupling. The ratio $\Lmax/\rnod$ was 
computed in \cite{pot:r0_SU3},
\bes \label{e_Lmaxrnod_old}
  \left.{\Lmax \over \rnod}\right|_{a=0} = 0.718(16)\,.
\ees
Now we know $\rnod$ closer to the continuum and can further investigate
the continuum extrapolation which lead to \eq{e_Lmaxrnod_old}. Details 
of the calculation, which are described in \cite{pot:r0_SU3}, are not repeated here. 
We only recall one important point. For the \SF of the pure gauge theory,
Symanzik's analysis implies that
lattice artifacts linear in the lattice spacing exist in general. 
They should vanish when the improvement coefficient $\ct$ (specific to the \SF)
is chosen properly. Since $\ct$ is know to 2-loop precision, it is interesting
to compare results for $\Lmax/\rnod$ obtained with  $\ct$ at 2-loop accuracy
to those using $\ct$ at  1-loop. 

\begin{figure}[tbp]
\hspace{0cm}
\vspace{-0.0cm}

\centerline{
\psfig{file=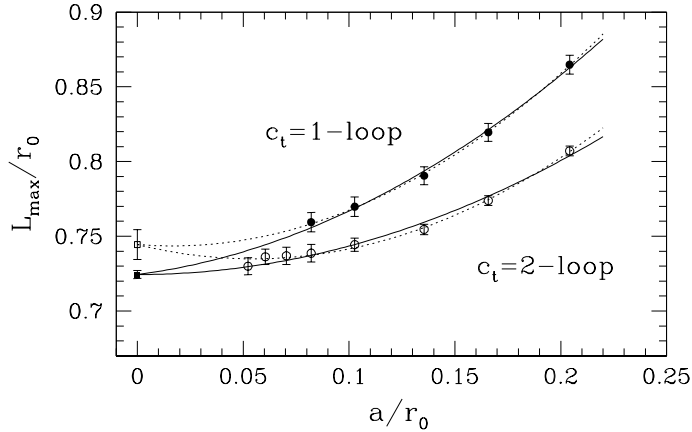,width=10cm}
}
\vspace{-0.0cm}
\caption{\footnotesize Fits to \protect\eq{e_cont_extr_ratio}. The fit
         represented by the full lines assumes $\rho_1^{(\rm 2-loop)}=0$,
         while for the dotted lines all coefficients are left as free parameters. 
\label{f_lmax_r0}}
\end{figure}
%
We obtained new data points by adding an entry $L/a=14\,,\;\beta=6.926(5)$ to Table 3
of \cite{pot:r0_SU3}\footnote{
We thank Jochen Heitger for performing the necessary simulation of the 
\SF.}
 and using $\rc/a$ and $\rc/\rnod$ from \sect{s_scales}. 
The resulting ratios
are shown in \fig{f_lmax_r0} together with fits of the form
\bes
  {\Lmax \over r_0} = \left.{\Lmax \over r_0} \right|_{a=0} 
                      + \rho_1^{(i)} {a \over r_0}
                      + \rho_2^{(i)} {a^2 \over r_0^2}\,, 
                      \quad i=\hbox{ 1-loop, 2-loop}\,.                
  \label{e_cont_extr_ratio}
\ees
If 2-loop approximation for $\ct$ is sufficient, 
we expect to have $\rho_1^{(\rm 2-loop)} \approx 0$. Our results are
compatible with this assumption (full lines, full square). 
On the other hand, when we leave $\rho_1^{(\rm 2-loop)}$ as a free
parameter, the best fit contains a noticeable linear term and
the continuum point $(\Lmax / r_0)_{a=0}$ differs significantly 
(dotted line and open square). 
Despite the 
precision of better than 1\% and the considerable range in the lattice spacing,
our data are not enough to decide between the two types of fits, which both 
have a good $\chi^2$.
As continuum limit, we quote
\bes
 \left.{\Lmax \over \rnod}\right|_{a=0} = 0.738(16)\,, \label{e_lmaxrnod}
\ees
covering the range allowed by both fits. Due to the
uncertainty in the proper extrapolation formula, the error is
not reduced compared to \eq{e_Lmaxrnod_old}. The small change 
in the central value 
hardly affects the value for the $\Lambda$-parameter.
We now have  $\Lambda_\msbar=0.586(48)/\rnod$ compared to
$\Lambda_\msbar=0.602(48)/\rnod$
of \cite{mbar:pap1}.

Our inability to improve the precision of the continuum limit
\eq{e_lmaxrnod} does not signal a problem with the standard theory 
of lattice artefacts. Rather we do have a difficult situation here where
both linear and quadratic lattice spacing effects are expected
and indeed both appear to be significant.


\section{Discussion}

In the pure gauge theory it was possible to compute the static potential
over a large range of distances by considering two sets
of lattices. Very small lattice spacings were used 
in the short distance range and 
larger ones for $r > 0.2\fm$. In a region of overlap the two sets of
computations could be matched (in the continuum limit). 
We find that the continuum force \fig{f_force} is a nice demonstration 
of the potential of lattice simulations for precision physics. 
The force
is in satisfactory agreement with perturbative predictions. A non-trivial 
point in this comparison is that the perturbative expression does not
involve any free parameter since the $\Lambda$-parameter in units of
$\rnod$ is known. Nevertheless there is a question of the truncation of
the perturbative series. Here we truncated the $\beta$-function
at 3-loop order. Other
possibilities and the stability of the perturbative prediction
will be discussed in a separate publication.

We have also investigated lattice spacing effects in some detail. 
First of all we have considered the question whether 
close to the continuum
the
leading lattice artefacts are quadratic in $a$ 
as commonly has been assumed. A positive answer to this
question could be given by formulating it in terms of the heavy quark effective 
theory for which Symanzik's discussion of $a$-effects is expected 
to apply. Also our numerical results are in full agreement
with $\rmO(a^2)$ lattice artifacts in potential differences. 

A numerically difficult situation is met when linear and quadratic
$a$-effects compete. This is the case in the ratio $\Lmax/\rnod$
considered in \sect{s_lmaxrnod}. Here the removal of linear $a$ effects is 
incomplete since the improvement coefficient $\ct$ is known only
perturbatively. Despite accurate numbers at finite 
lattice spacing our continuum estimate  \eq{e_lmaxrnod}
has a relatively large uncertainty. In order to avoid such a situation
in simulations of full QCD, non-perturbative $\Oa$-improvement
\cite{impr:pap2} should be applied or one should use formulations
with exact chiral symmetry where linear terms are
absent without any necessity of tuning 
improvement coefficients~\cite{reviews:exact_chi_sym}.

We finally point out that at our level
of precision the observed lattice spacing effects are
in complete agreement with the standard theory of Symanzik. 
Our precision does however not quite reach the one 
achieved in 2 dimensions, where
indications for deviations from the standard picture
have been found \cite{sigma:unexpected}. 
\vspace{0.8cm}

\noindent
{\bf Acknowledgment}\\
\noindent
We would like to thank Marco Guagnelli and Hartmut Wittig for 
the data generated in an earlier project \cite{pot:r0_SU3} of
the ALPHA collaboration as well as for discussions and
comments on our manuscript. It is also
a pleasure to thank Francesco Knechtli for his help
and the SU(2) code which was the starting point for our program. Our new 
simulations were performed at the Konrad-Zuse-Zentrum f\"ur Informationstechnik
Berlin (ZIB).
We thank this center for granting CPU-resources to this project
and Hinnerk St\"uben for assistance.


\appendix
\section{Computation of potential and force \label{app_a}}
We computed Wilson loop correlation matrices using smearing 
exactly as in \cite{pot:r0_SU3}. Three different smearing levels
$n_l, l=0,1,2$
were used, constructing a $3\times 3$ matrix
for each value of the loop size $r,t$ with $a \leq r \leq r_{\rm max}$ and
$a \leq t \leq t_{\rm max}$. Our choices 
for the different parameters are listed in \tab{parameters}.
The numerical values for $n_l$, defined as in \cite{pot:r0_SU3},
are in general smaller than in that reference since we want to compute the
potential for shorter distances. For the same reason only 40 to 75 
measurements were sufficient
to get satisfactory precision. With relatively modest computational effort
we could extend the calculations
to $a \approx 0.025 \fm$ where a $64^4$ lattice had to be simulated. 
Several iterations 
of a hybrid overrelaxation algorithm with $N_{\rm or}$ overrelaxation
sweeps per heatbath sweep were performed to separate the field configurations
of our MC sample.

\begin{table}\caption{\label{parameters}\footnotesize{Simulation 
parameters. }}
\begin{center}
\begin{tabular}{cccccc}
\hline
$L/a$ & $\beta$ & $n_{0},~n_{1},~n_{2}~$ & $N_{\rm or}$ & $r_{\rm max}/a$ & $t_{\rm max}/a$  \\
\hline
40    &  6.57 & 0,~77,~153 & 18 & 7~~  & 11  \\
48    &  6.69 & 0,~53,~106 & 22 & 10 & 12  \\
56    & 6.81  & 0,~72,~144 & 25 & 12 & 14  \\
64    & 6.92  & 0,~94,~188 & 29 & 12 & 16  \\ 
\hline
\end{tabular}
\end{center}
\end{table}


The potential $V(r)$  was extracted 
from the correlation matrices
following the procedure of  \cite{pot:r0_SU3}. 
As in this reference, we found that for $t>t_0$ excited
state contaminations are negligible when $\exp[-t_0(V_1(r)-V_0(r)]<0.3$ is
satisfied. Since the gap $V_1(r)-V_0(r)$ grows towards small $r$
from $\approx 3 /\rnod$ at $r=\rnod$ to  $\approx 7 /\rnod$ at $r=0.1 \rnod$,
the short distance region is easiest in this respect.
All errors quoted in this paper 
were computed by jacknife binning. Force and potential are listed in 
Tables 4-7. 
\vspace{0.4cm}
  \begin{table}[ht] 
\caption{\label{t_force2}\footnotesize{Force and potential in the
short distance region. }}  
\begin{center}
  \begin{tabular}{ c  c  c  c  c }
  \hline 
 $\beta$  & $\rI/a$  & $a^2\,F(\rI)$  &  $d_{\rm I}/a$ &  $a\,V_{\rm I}(d_{\rm I})$     \\  
  \hline 
 $ 6.57 $ & $  $ & $   $ & $ 1.855 $ & $ 0.457898 ( 71 )$ \\
 $  $ & $ 2.277 $ & $ 0.056623 ( 65 ) $ & $ 2.889 $ & $ 0.51452 ( 12 )$ \\
 $  $ & $ 3.312 $ & $ 0.033391 ( 97 ) $ & $ 3.922 $ & $ 0.54795 ( 21 )$ \\
 $  $ & $ 4.359 $ & $ 0.023752 ( 94 ) $ & $ 4.942 $ & $ 0.57170 ( 28 )$ \\
 $  $ & $ 5.393 $ & $ 0.01904 ( 12 ) $ & $ 5.954  $ & $ 0.59074 ( 35 )$ \\
 $  $ & $ 6.414 $ & $ 0.01629 ( 13 ) $ & $ 6.962  $ & $ 0.60703 ( 41 )$ \\[1ex]
 $ 6.69 $ & $  $ & $   $ & $ 1.855 $ & $ 0.43918 ( 41 )$ \\
 $  $ & $ 2.277 $ & $ 0.052308 ( 38 ) $ & $ 2.889 $ & $ 0.491487 ( 72 )$ \\
 $  $ & $ 3.312 $ & $ 0.030174 ( 66 ) $ & $ 3.922 $ & $ 0.52162 ( 13 )$ \\
 $  $ & $ 4.359 $ & $ 0.021054 ( 75 ) $ & $ 4.942 $ & $ 0.54267 ( 18 )$ \\
 $  $ & $ 5.393 $ & $ 0.016439 ( 72 ) $ & $ 5.954 $ & $ 0.55911 ( 23 )$ \\
 $  $ & $ 6.414 $ & $ 0.013728 ( 82 ) $ & $ 6.962 $ & $ 0.57284 ( 28 )$ \\
 $  $ & $ 7.428 $ & $ 0.012036 ( 96 ) $ & $ 7.967 $ & $ 0.58487 ( 35 )$ \\
 $  $ & $ 8.438 $ & $ 0.010869 ( 82 ) $ & $ 8.971  $ & $ 0.59574 ( 41 )$ \\
 $  $ & $ 9.445 $ & $ 0.010123 ( 97 ) $ & $ 9.974 $ & $ 0.60587 ( 46 )$ \\[1ex]
 $ 6.81 $ & $  $ & $   $ & $ 1.855 $ & $ 0.422617 ( 22 )$ \\
 $  $ & $ 2.277 $ & $ 0.048833 ( 26 ) $ & $ 2.889 $ & $ 0.471411 ( 47 )$ \\
 $  $ & $ 3.312 $ & $ 0.027650 ( 32 ) $ & $ 3.922  $ & $ 0.499062 ( 66 )$ \\
 $  $ & $ 4.359 $ & $ 0.018860 ( 31 ) $ & $ 4.942 $ & $ 0.517921 ( 90 )$ \\
 $  $ & $ 5.393 $ & $ 0.014471 ( 32 ) $ & $ 5.954 $ & $ 0.53239 ( 11 )$ \\
 $  $ & $ 6.414 $ & $ 0.011870 ( 40 ) $ & $ 6.962 $ & $ 0.54426 ( 13 )$ \\
 $  $ & $ 7.428 $ & $ 0.010163 ( 54 ) $ & $ 7.967 $ & $ 0.55437 ( 18 )$ \\
 $  $ & $ 8.438 $ & $ 0.009072 ( 56 ) $ & $ 8.971 $ & $ 0.56344 ( 21 )$ \\
 $  $ & $ 9.445 $ & $ 0.008267 ( 52 ) $ & $ 9.974 $ & $ 0.57171 ( 24 )$ \\
 $  $ & $ 10.451$ & $ 0.007701 ( 58 ) $ & $ 10.977 $ & $ 0.57941 ( 27 )$ \\
 $  $ & $ 11.455$ & $ 0.007232 ( 62 ) $ & $ 11.979 $ & $ 0.58664 ( 30 )$ \\
  \hline 
  \end{tabular} 
\end{center}
  \end{table} 
  \begin{table}[ht] 
\caption{\label{t_force2a}\footnotesize{Force and potential in the
short distance region. }}  
\begin{center}
  \begin{tabular}{ c  c  c  c  c }
  \hline 
 $\beta$  & $\rI/a$  & $a^2\,F(\rI)$  &  $d_{\rm I}/a$ &  $a\,V_{\rm I}(d_{\rm I})$     \\  
  \hline 
 $ 6.92 $ & $  $ & $   $ & $ 1.855 $ & $ 0.408642 ( 19 )$ \\
 $  $ & $ 2.277 $ & $ 0.046081 ( 19 ) $ & $ 2.889 $ & $ 0.454723 ( 34 )$ \\
 $  $ & $ 3.312 $ & $ 0.025696 ( 24 ) $ & $ 3.922 $ & $ 0.480418 ( 47 )$ \\
 $  $ & $ 4.359 $ & $ 0.017266 ( 29 ) $ & $ 4.942 $ & $ 0.497662 ( 75 )$ \\
 $  $ & $ 5.393 $ & $ 0.012969 ( 33 ) $ & $ 5.954 $ & $ 0.510631 ( 88 )$ \\
 $  $ & $ 6.414 $ & $ 0.010412 ( 39 ) $ & $ 6.962 $ & $ 0.52104 ( 11 )$ \\
 $  $ & $ 7.428 $ & $ 0.008855 ( 37 ) $ & $ 7.967 $ & $ 0.52990 ( 13 )$ \\
 $  $ & $ 8.438 $ & $ 0.007755 ( 43 ) $ & $ 8.971  $ & $ 0.53765 ( 16 )$ \\
 $  $ & $ 9.445 $ & $ 0.006974 ( 51 ) $ & $ 9.974 $ & $ 0.54463 ( 19 )$ \\
 $  $ & $ 10.451$ & $ 0.006402 ( 53 ) $ & $ 10.977 $ & $ 0.55103 ( 23 )$ \\
 $  $ & $ 11.455$ & $ 0.006022 ( 53 ) $ & $ 11.979 $ & $ 0.55705 ( 26 )$ \\
  \hline 
  \end{tabular} 
\end{center}
  \end{table} 


  \begin{table}[ht] 
\caption{\label{t_force1}\footnotesize{Force and potential
from the data of Guagnelli et al. \protect\cite{pot:r0_SU3}}}  
\begin{center}
\begin{tabular}{ c  c  c  c  c }
  \hline 
 $\beta$  & $\rI/a$  & $a^2\,F(\rI)$  &  $d_{\rm I}/a$ &  $a\,V_{\rm I}(d_{\rm I})$     \\ 
  \hline 
 $ 5.95 $ & $  $ & $   $ & $ 1.855 $ & $ 0.61794 ( 16 )$ \\
 $  $ & $ 2.277 $ & $ 0.11212 ( 20 ) $ & $ 2.889 $ & $ 0.73006 ( 33 )$ \\
 $  $ & $ 3.312 $ & $ 0.08319 ( 27 ) $ & $ 3.922 $ & $ 0.81325 ( 55 )$ \\
 $  $ & $ 4.359 $ & $ 0.07164 ( 36 ) $ & $ 4.942 $ & $ 0.88489 ( 84 )$ \\
 $  $ & $ 5.393 $ & $ 0.06613 ( 48 ) $ & $ 5.954 $ & $ 0.9510 ( 12 )$ \\
 $  $ & $ 6.414 $ & $ 0.06296 ( 56 ) $ & $ 6.962 $ & $ 1.0140 ( 17 )$ \\
 $  $ & $ 7.428 $ & $ 0.0606 ( 12 ) $ & $ 7.967 $ & $ 1.0728 ( 30 )$ \\[1ex]
 $ 6.07 $ & $  $ & $   $ & $ 1.855 $ & $ 0.571729 ( 97 )$ \\
 $  $ & $ 2.277 $ & $ 0.09211 ( 11 ) $ & $ 2.889 $ & $ 0.66384 ( 19 )$ \\
 $  $ & $ 3.312 $ & $ 0.06427 ( 13 ) $ & $ 3.922 $ & $ 0.72811 ( 30 )$ \\
 $  $ & $ 4.359 $ & $ 0.05301 ( 26 ) $ & $ 4.942 $ & $ 0.78116 ( 55 )$ \\
 $  $ & $ 5.393 $ & $ 0.04771 ( 28 ) $ & $ 5.954 $ & $ 0.82887 ( 72 )$ \\
 $  $ & $ 6.414 $ & $ 0.04468 ( 27 ) $ & $ 6.962 $ & $ 0.87355 ( 89 )$ \\
 $  $ & $ 7.428 $ & $ 0.04262 ( 38 ) $ & $ 7.967 $ & $ 0.9162 ( 11 )$ \\
 $  $ & $ 8.438 $ & $ 0.04215 ( 45 ) $ & $ 8.971 $ & $ 0.9583 ( 13 )$ \\
 $  $ & $ 9.445 $ & $ 0.04087 ( 73 ) $ & $ 9.974 $ & $ 0.9992 ( 17 )$ \\[1ex]
 $ 6.2 $ & $  $ & $   $ & $ 1.855 $ & $ 0.533457 ( 82 )$ \\
 $  $ & $ 2.277 $ & $ 0.07804 ( 11 ) $ & $ 2.889 $ & $ 0.61145 ( 17 )$ \\
 $  $ & $ 3.312 $ & $ 0.05135 ( 14 ) $ & $ 3.922 $ & $ 0.66279 ( 28 )$ \\
 $  $ & $ 4.359 $ & $ 0.04054 ( 13 ) $ & $ 4.942 $ & $ 0.70333 ( 38 )$ \\
 $  $ & $ 5.393 $ & $ 0.03511 ( 20 ) $ & $ 5.954 $ & $ 0.73844 ( 54 )$ \\
 $  $ & $ 6.414 $ & $ 0.03238 ( 20 ) $ & $ 6.962 $ & $ 0.77082 ( 69 )$ \\
 $  $ & $ 7.428 $ & $ 0.03018 ( 25 ) $ & $ 7.967 $ & $ 0.80100 ( 85 )$ \\
 $  $ & $ 8.438 $ & $ 0.02884 ( 26 ) $ & $ 8.971 $ & $ 0.8298 ( 10 )$ \\
 $  $ & $ 9.445 $ & $ 0.02813 ( 27 ) $ & $ 9.974 $ & $ 0.8580 ( 12 )$ \\
 $  $ & $ 10.451 $ & $ 0.02766 ( 30 ) $ & $ 10.977 $ & $ 0.8856 ( 14 )$ \\
 $  $ & $ 11.455 $ & $ 0.02752 ( 34 ) $ & $ 11.979 $ & $ 0.9131 ( 16 )$ \\
  \hline 
  \end{tabular} 
\end{center}
  \end{table} 
  \begin{table}[ht] 
\caption{\label{t_force1a}\footnotesize{Force and potential
from the data of Guagnelli et al. \protect\cite{pot:r0_SU3}}}  
\begin{center}
\begin{tabular}{ c  c  c  c  c }
  \hline 
 $\beta$  & $\rI/a$  & $a^2\,F(\rI)$  &  $d_{\rm I}/a$ &  $a\,V_{\rm I}(d_{\rm I})$     \\ 
  \hline 
 $ 6.4 $ & $  $ & $   $ & $ 1.855 $ & $ 0.488379 ( 48 )$ \\
 $  $ & $ 2.277 $ & $ 0.064318 ( 51 ) $ & $ 2.889 $ & $ 0.552697 ( 81 )$ \\
 $  $ & $ 3.312 $ & $ 0.039580 ( 70 ) $ & $ 3.922 $ & $ 0.59228 ( 13 )$ \\
 $  $ & $ 4.359 $ & $ 0.029360 ( 79 ) $ & $ 4.942 $ & $ 0.62164 ( 19 )$ \\
 $  $ & $ 5.393 $ & $ 0.024367 ( 84 ) $ & $ 5.954 $ & $ 0.64600 ( 24 )$ \\
 $  $ & $ 6.414 $ & $ 0.02145 ( 12 ) $ & $ 6.962 $ & $ 0.66744 ( 40 )$ \\
 $  $ & $ 7.428 $ & $ 0.01939 ( 16 ) $ & $ 7.967 $ & $ 0.68683 ( 50 )$ \\
 $  $ & $ 8.438 $ & $ 0.01819 ( 18 ) $ & $ 8.971 $ & $ 0.70502 ( 66 )$ \\
 $  $ & $ 9.445 $ & $ 0.01757 ( 17 ) $ & $ 9.974 $ & $ 0.72258 ( 79 )$ \\
 $  $ & $ 10.451 $ & $ 0.01677 ( 17 ) $ & $ 10.977 $ & $ 0.73936 ( 84 )$ \\
 $  $ & $ 11.455 $ & $ 0.01651 ( 15 ) $ & $ 11.979 $ & $ 0.75586 ( 94 )$ \\
 $  $ & $ 12.459 $ & $ 0.01609 ( 17 ) $ & $ 12.980 $ & $ 0.7720 ( 11 )$ \\
 $  $ & $ 13.462 $ & $ 0.01616 ( 29 ) $ & $ 13.982 $ & $ 0.7881 ( 11 )$ \\
 $  $ & $ 14.465 $ & $ 0.01564 ( 18 ) $ & $ 14.983 $ & $ 0.8038 ( 12 )$ \\
 $  $ & $ 15.467 $ & $ 0.01513 ( 39 ) $ & $ 15.984 $ & $ 0.8189 ( 14 )$ \\
  \hline 
  \end{tabular} 
\end{center}
  \end{table} 


\noindent
{\em Interpolation.}

Solving \eq{e_rnod} and evaluating force and potential 
at distances $r=x\rc$ for given $x$, requires their interpolation. While
this can in principle be done in many ways, it is advantageous
to use physics motivated interpolation formulae.

For the force we followed \cite{pot:r0} choosing the interpolation function
\be
F(r)=f_{1}+f_{2}r^{-2},
\ee
between the two neighboring points. 
The systematic error arising from the 
interpolation was estimated adding a term $f_{3}r^{-4}$ and taking a third point. 
The difference between the two interpolations was added (linearly) to
the statistical uncertainty. We observed that at least for $r\grtsim 0.4\rnod$ the 
interpolation error is smaller than the statistical uncertainty. For small distances 
the systematic error increases and can be of the order of the statistical one
or even bigger.

The potential is interpolated by the corresponding 
ansatz
\be
\VI(r)=v_{1}+v_{2}r+v_{3}r^{-1}.
\ee
Here two points are chosen such that the desired value of $r$ is in between.
For the choice of a neighboring third point one has two possibilities, leading to 
two results.
Their difference was taken as the interpolation error. Also in this case 
the systematic 
errors are larger than the statistical ones at short distances  while
at large distances the situation is reversed.


\section{The evaluation of the 3-dimensional lattice propagator in 
coordinate space
\label{app_d}}
An efficient method to calculate the lattice propagator in 
coordinate space is proposed in \cite{pert:LW95b}.  
It is based on a recursion relation which allows to express the propagator 
as linear function of its values near the origin. In this paper, 
the 4-dimensional case is discussed; it is straitghforward to apply 
the same method in 3 dimensions. We use lattice units, $a=1$, in this appendix.

Starting from the Laplace equation for the Green function,
\be\label{green}
-\triangle G(\vecx)=\left\{\begin{array}{ll}
1 & \textrm{if $\vecx=0$}\\
0 & \textrm{otherwise}
\end{array}\right.,
\ee
with 
\be
\triangle f(\vecx)=  \sum_{j=1}^{3}\left[f(\vecx+\hat{j}) -2f(\vecx) + f(\vecx-\hat{j})
                                   \right],
\ee
and using other general features of the lattice propagator, one obtains ($j=1,2,3$)
\be\label{recursion}
G(\vecx+\hat{j})=G(\vecx-\hat{j})+2 {x_{j}\over\rho}\sum_{i=1}^{3}[G(\vecx)-G(\vecx-\hat{i})]
\,, 
\ee
for $\rho=\sum_{j=1}^{3}x_{j}\neq 0$.

Because of isotropy, \eq{recursion} 
can be restricted to the points $\vecx$ with $x_{1}\geq x_{2} \geq x_{3} \geq 0$. 
In this region \eq{recursion} can be used as recursion relation to express $G(\vecx)$ 
as a linear combination of
$
G(0,0,0),\;G(1,0,0),\; G(1,1,0),\; G(1,1,1).
$
These four values at the corners of the unit cube are not independent: 
from \eq{green} at $\vecx=0$
\be\label{relation1}
G(0,0,0)-G(1,0,0)=\frac{1}{6}
\ee
follows directly. Another relation can be deduced in the following way:
first one observes that \eq{recursion} becomes one-dimensional 
along the lattice axes. Defining
\be
g_{1}(n)=G(n,0,0),
\,\;
g_{2}(n)=G(n,1,0),
\,\;
g_{3}(n)=G(n,1,1)
\ee
and using the lattice symmetries, we find
\bes
g_{1}(n+1)&=&6g_{1}(n)-4g_{2}(n)-g_{1}(n-1)\,, \\
g_{2}(n+1)&=&\frac{2n}{n+1}\left[3g_{2}(n)-g_{1}(n)-g_{3}(n)\right]-\frac{n-1}{n+1}g_{2}(n-1)\,,
\\
g_{3}(n+1)&=&\frac{2n}{n+2}\left[3g_{3}(n)-2g_{2}(n)\right]-\frac{n-2}{n+2}g_{3}(n-1)\,.
\ee
Next one notices that
\bes\label{invariant}
k(n)&=& (n-1)g_{1}(n)+2ng_{2}(n)+(n+1)g_{3}(n)-ng_{1}(n-1)- \\
    & & 2(n-1)g_{2}(n-1)-(n-2)g_{3}(n-1) \nonumber
\ees
is an invariant of the recursion relation, that is
$
k(n+1)=k(n)
$
for $n\geq 1$.
The value of $k$ is worked out in the limit $n\rightarrow\infty$, where\\
$$
g_{j}(n)=\frac{1}{4\pi n}+O(1/n^{2}), \quad \textrm{for} \quad j=1,2,3\,.
$$
This consideration yields $k(n)=0$ for $n\geq 1$.
Setting $n=1$ in \eq{invariant} one obtains\\
\be\label{relation2}
3G(1,1,0)+2G(1,1,1)-G(0,0,0)=0.
\ee
Thus from the four initial values two can be eliminated and the propagator is obtained in the form\\
\be\label{principal}
G(\vecx)=r_{1}(\vecx)G(0,0,0)+r_{2}(\vecx)G(1,1,0)+r_{3}(\vecx).
\ee
The coefficients $r_{1},r_{2},r_{3}$ are rational numbers and 
can be evaluated recursively from \eq{recursion}.

From the numerical point of view, we are then left with the task to 
accurately compute 
$G(0,0,0)$ and $G(1,1,0)$. This can be done with the procedure discussed 
in  \cite{pert:LW95b}, giving
$$
G(0,0,0)=0.2527310098586630030260020266135701299...,
$$
$$
G(1,1,0)=0.0551914336877373170165449460300639378...
$$


%
   \bibliography{pap1}        
   \bibliographystyle{h-elsevier}   
\end{document}